\documentclass[pre,showpacs,showkeys,10pt,superscriptaddress]{revtex4}

\usepackage{epsfig}
\usepackage{graphicx}
\usepackage{amsmath}
\usepackage{amssymb}
\usepackage{amsfonts}

\begin{document}

\title{Collisionless plasma expansion in the presence of a dipole magnetic field}
\author{H. B. Nersisyan}
\email{hrachya@irphe.am}
\affiliation{Division of Theoretical Physics, Institute of Radiophysics and
Electronics, Alikhanian Brothers Street 1, 378410 Ashtarak, Armenia}
\affiliation{Centre of Strong Fields Physics, Yerevan State University, Alex Manoogian
str. 1, 375025 Yerevan, Armenia}
\author{D. A. Osipyan}
\affiliation{Division of Theoretical Physics, Institute of Radiophysics and
Electronics, Alikhanian Brothers Street 1, 378410 Ashtarak, Armenia}

\begin{abstract}
The collisionless interaction of an expanding high--energy plasma cloud
with a magnetized background plasma in the presence of a dipole magnetic
field is examined in the framework of a 2D3V hybrid (kinetic ions and
massless fluid electrons) model. The retardation of the plasma cloud and
the dynamics of the perturbed electromagnetic fields and the background
plasma are studied for high Alfv\'{e}n--Mach numbers using the
particle--in--cell method. It is shown that the plasma cloud expands
excluding the ambient magnetic field and the background plasma to form a
diamagnetic cavity which is accompanied by the generation of a collisionless
shock wave. The energy exchange between the plasma cloud and the background
plasma is also studied and qualitative agreement with the analytical model
suggested previously is obtained.
\end{abstract}


\maketitle

\section{Introduction}
\label{sec:intr}

When a high--energy density plasma is expanded into a magnetized
background plasma, the injected plasma expands and excludes the
background plasma and magnetic field to form a diamagnetic cavity.
This process has been observed in a number of experiments in space
and the laboratory (see, e.g., Ref.~\cite{zak03} and references
therein). The similar processes are responsible
for the formation of the heliopause, where solar wind drags or
convects magnetic fields from the Sun, which causes the deflection
of both the interstellar medium and cosmic rays. Other examples
of naturally occurring inflated or stretched magnetic fields include
coronal mass ejections and the formation of the Earth's magnetotail
\cite{win05,win00,win98}. Starting in the 1960s, many experiments
of the collisionless phenomena of exploding plasmas in
space and laboratory were performed or planned. Dense, high--energy
plasmas can also be formed by laser--irradiation of a small target
embedded in a gas in an external magnetic field \cite{rip93}. The
background gas can either be pre--ionized or be ionized by the laser.
Such experiments show that, in addition to expansion, the target
plasma is also subject to flute instabilities on its surface as it
interacts with the background field and embedded plasma. The
diamagnetic cavities are expected to occur also in supernova
explosions \cite{dra02}. The physics of the plasma expansion and
evolution has been investigated in detail numerically, using a
variety of techniques, such as full particle \cite{win88}, hybrid
(particle ions, fluid electron) \cite{bre87} and magnetohydrodynamics
(MHD) \cite{hub92} codes. The properties of the expansion phase in
the presence of either a stationary \cite{gis89} or a flowing
background plasma \cite{win89}, as well as the details of unstable
modes on the surface of the expanding plasma \cite{win89,hub90,bre92},
have been well studied.

Typical experimental parameters are such that the ions are collisionless
but the situation for the electrons ranges from collisionless to
collision--dominated. In this paper the dynamics of the expanding plasma cloud is
studied for the completely collisionless regimes, which, in particular,
are realized in many astrophysical processes (see, e.g., \cite{zak03} and
references therein). An example is the problem of the collisionless retardation
of the supernovae remnants which was first formulated by Oort \cite{oor46}
and subsequently analyzed in detail by Shklovskii \cite{shk76}.

In recent years, the basic concept of the plasma expansion has been extended
to consider a large magnetic bubble, which can be formed using a small magnetic
coil and plasma source attached to a spacecraft, to efficiently inflate the
bubble to a large cross--sectional area \cite{win00,win03}. A net force would
be exerted on the spacecraft due to the deflection of the solar wind around
the bubble. In this case, the plasma is continuously injected in the presence
of a dipole or dipole--like magnetic field. Theoretical arguments \cite{par03} and calculations
(see, e.g., Refs.~\cite{win00,tan07,gar08} and references therein) suggest that the
magnetic field of the dipole can be expanded with the plasma, so that the
magnitude of the field falls off much slower with distance $r$ from the dipole,
namely as $r^{-s}$, with $s \sim 1,2$ in certain directions at least, rather
than $r^{-3}$ (for a bare dipole), allowing a large bubble to form. Laboratory
experiments \cite{win01,win02} have provided some evidence for the slow falloff
of the field from the source. However, the nature of the plasma and magnetic
field expansion in this configuration is not completely understood, and how
it compares with the more common picture of diamagnetic cavity formation has
not been addressed to date. References~\cite{zak03,win05,tan07} and
the papers cited therein conducted a systematic study in a two--dimensional
(2D) geometry and only for initial phase of expansion \cite{win05}, and in a
three--dimensions (3D) but mainly of the characteristics of the magnetic field
inflation \cite{tan07}. Special attention has been paid to the MHD analysis
of the expanding plasma behavior in a dipole field in a vacuum \cite{mur01},
i.e. in the absence of the background plasma.

In this paper, in order to provide further insight into the underlying physics
we have performed systematic particle--in--cell (PIC) 2D3V hybrid simulations of
the plasma expansion in a magnetized background plasma in the presence of a
dipole magnetic field. The basic parameters of the problem are introduced in
Sec.~\ref{sec:1}. The numerical method and model used are discussed in
Sec.~\ref{sec:2}. We perform a number of simulations assuming hydrogen cloud
and hydrogen background plasmas and the results are presented and discussed in
Sec.~\ref{sec:3}. Finally, we state the conclusions in Sec.~\ref{sec:conc}.

\section{Basic parameters for the plasma expansion}
\label{sec:1}

The plasma expansion process is characterized by the magnetic ($R_{m}$) and
hydrodynamic ($R_{g}$) retardation lengths. $R_{m}$ is obtained by equating
the initial kinetic energy $W_{0}$ of an initially spherical plasma cloud to
the energy of the magnetic field that it pushes out in expanding to the radius
$R_{m}$ \cite{rai63}, i.e., $R_{m}=(6W_{0}/H_{0}^{2})^{1/3}$. Here $H_{0}$ is
the strength of the unperturbed magnetic field. As the plasma expands, it draws
the background plasma into a combined motion. Along with this, the mass of
expelled plasma increases. The radius of the sphere within which the mass of
the plasma cloud and that of the background plasma drawn into the combined
motion become equal is referred to as the hydrodynamic retardation length:
$R_{g}=(3M/4\pi n_{b}m_{b})^{1/3}$, where $n_{b}$ and $m_{b}$ are the density
and mass of the background plasma ions \cite{shk76} and $M$ is the mass of
the expanding plasma. The smaller of the lengths, $R_{m}$ or $R_{g}$, determines
the predominant mechanism for the retardation of the plasma cloud--magnetic or
hydrodynamic. Of course $R_{m}$ and $R_{g}$ account for the ideal characteristics
of the retardation process, therefore some high--energy particles may penetrate
beyond the stopping length.
The relation $R_{m}/R_{g}\sim M_{A}^{2/3}$, where $M_{A}=u/v_{A}$
is the Alfv\'{e}n--Mach number ($u\sim (W_{0}/M)^{1/2}$ is the initial
expansion velocity of the plasma) and $v_{A}=H_{0}/\sqrt{4\pi n_{b}m_{b}}$ is
the Alfv\'{e}n velocity in the background plasma, implies that for $M_{A}\ll 1$
the cloud loses energy as a result of the deformation and displacement of the
magnetic field, while for $M_{A}\gg 1$ the retardation is caused by the
interaction with the background plasma \cite{zak88,osi03} (see also
Ref.~\cite{zak03}). In this paper we consider only the second (hydrodynamic)
regime of collisionless expansion with $M_{A}\gg 1$. In this case the retardation
length is given by $\ell_{s} \simeq R_{g}$.

An analysis shows that the hydrodynamic retardation can only be ensured by
a collisionless laminar (or turbulent) mechanism \cite{zak88} associated
with the generation of vortical electric fields $E_{i}$ in the front of
the expanding plasma or by a collisional mechanism owing to pairwise
collisions of the cloud ions with the ions and electrons of the background plasma.
The ions of the expanding plasma transfer their energy to the ions and
electrons of the background plasma in multiple ion--ion or ion--electron Coulomb
collisions with a mean free paths $\lambda _{ii}$ and $\lambda _{ei}\sim (m_{e}/m_{c})%
\lambda _{ii}$, respectively \cite{tru65,koo72}. Here $m_{e}$ and $m_{c}$ are the
electron and cloud ion masses, respectively. In this paper we assume a completely
collisionless regimes when $\lambda_{ii}\gg \lambda_{ei}>\ell_{s}$. Strictly
speaking, the hydrodynamic description is inapplicable under these conditions.
However, as shown in Ref.~\cite{rai95} the hydrodynamic approach, when considering
collisionless plasma expansion into the magnetized, ionized medium, can give
quite reasonable qualitative and even satisfactory quantitative results. The
physical reason underlying this situation is the small ion cyclotron radius $a_{L}$
in comparison with a characteristic flow scale $\ell_{s}$, with the ion cyclotron
radius playing the role of particle mean free path.

The first group of the collisionless interaction mechanisms are collective
turbulent mechanisms (anomalous viscosity, anomalous resistivity) during
the development of ion--ion or electron--ion beam instabilities \cite{win84}.
The condition for excitation of the ion--ion instability has the form
$u^{2}\leqslant v_{A}^{2}+2c_{s}^{2}$ (see, e.g., Ref.~\cite{pap71}) or
$M_{A}^{2}\leqslant 1+2c_{s}^{2}/v_{A}^{2}$, where $c_{s}=\sqrt{T_{e}/m_{b}}$
is the ion sound speed in the background plasma. In many typical situations
$v_{A}\gtrsim c_{s}$ and $M_{A}\lesssim 2$. Thus, according to the criteria
mentioned above at high Alfv\'{e}n--Mach numbers $M_{A}\gg 1$ the retardation
of the expanding plasma cannot be caused by the turbulent mechanisms.

The second group is the collisionless laminar retardation mechanism associated
with the generation of vortical electric fields. It is known that the role
of vortical electric fields becomes predominant as $M_{A}$ increases, since
$E_{i}/E_{p}\sim M_{A}^{2}$ \cite{zak88}, where $E_{p}$ is the polarization
electric field which arises as a result of the drop of the hydrodynamic and
magnetic pressures at the front of the plasma cloud. A model for energy exchange
between the cloud and the background plasma due to the combined effect of the
gyromotion of the ions and the generation of vortical electric fields when
$M_{A}>1$ (so--called magnetolaminar mechanism (MLM)) has been proposed in
Ref.~\cite{gol78}. Analytic solutions for the initial expansion phase, when
only a vortical electric field $E_{i}$ develops, showed that the fraction of
energy transferred from the cloud to the background plasma is proportional to
$\delta =(R_{g}/a_{L})^{2}$ (MLM interaction parameter), where $a_{L}$ is the
cyclotron radius of the cloud ions.

In the framework of the ideal MHD approximation for the description of the plasma
expansion into a vacuum in the presence of dipole magnetic field, another
energetic parameter $\kappa $ was defined which is given by $\kappa =W_{0}/W_{M}$
\cite{nik93}, where $W_{M}$ is the total magnetic energy of the dipole beyond
the spherical radius $r_{d}$ ($W_{M}=p^{2}/3r_{d}^{3}$), $r_{d}$ is the distance
from the dipole to the center of the plasma cloud and $p$ is the magnetic moment
magnitude. Nikitin et al. introduced critical value $\kappa_{c}$ for
a different plasma location \cite{nik93}. In the case when $\kappa <\kappa_{c}$,
a substantial plasma retardation will occur in all directions from the cloud center
location ("quasi--capture" mode), meanwhile the plasma will not be captured by
an ambient magnetic field when $\kappa >\kappa_{c}$ ("rupture" mode). The critical
value varies between $0.1$ and $0.4$. The latter value is realized when the plasma
cloud is located at the dipole axis. In this paper we consider only the second
regime of the plasma expansion.

\section{Collisionless hybrid simulation}
\label{sec:2}

The hybrid model is used to study the collisionless plasma expansion
in a magnetic field. This model describes the ions by means of the
velocity distribution function $f(\mathbf{r},\mathbf{v},t)$, the
electrons being considered hydrodynamically \cite{hae81,ler84}. The
typical scale on which the macroscopic parameters (the plasma density,
magnetic field) vary is the ion cyclotron radius $a_{L}$.
The electron cyclotron radius $a_{Le}$ is much less than $a_{L}$ due
to the small electron mass. Of course, the electron distribution
function changes sharply on a small scale $a_{Le}$, but we do not
consider such details. If we are interested in the behavior of those
macroscopic parameters that vary on a scale $a_{L}$, it is sufficient
to consider only the motion of the electron small--scale gyromotion
centers. In this case one can consider the electrons as a fluid and
describe their motions hydrodynamically.

All of the estimates show that the plasma under the conditions in
question is essentially quasineutral everywhere, although there are
thin space charge regions. These regions are located near the
plasma cloud boundary as well as near the collisionless shock in the
background plasma. The scales of these regions are essentially less
than $a_{L}$, and therefore, they are not considered here. Then the
densities of ions $n_{\alpha}$ and electrons $n_{e}$ in quasineutral
plasma satisfy the condition $n_{e}=\sum_{\alpha}Z_{\alpha}n_{\alpha}$,
where $Z_{\alpha}e$ is the charge of the ion from plasma species $\alpha$.
Let the cloud ions have the index $\alpha =c$, whereas the index of
the background ions is $\alpha =b$. The ion distribution functions
$f_{\alpha}$ are governed by the Vlasov kinetic equation
\begin{equation}
\frac{\partial f_{\alpha}}{\partial t}+{\mathbf{v}}\cdot \frac{\partial f_{\alpha}}{%
\partial {\mathbf{r}}}+\frac{Z_{\alpha}e}{m_{\alpha}}\left( {\mathbf{E}}+\frac{1}{c}\left[
{\mathbf{v}}\times {\mathbf{H}}\right] \right) \cdot \frac{\partial f_{\alpha}}{%
\partial {\mathbf{v}}}=0 .
\label{eq:4}
\end{equation}%
The electric and magnetic fields satisfy the Maxwell equations
\begin{equation}
\boldsymbol{\nabla }\times {\mathbf{H}}=\frac{4\pi n_{e}e}{c}\left( \mathbf{v}_{i}-%
\mathbf{v}_{e}\right) ,\quad \boldsymbol{\nabla }\times \mathbf{E}=-\frac{1}{c}%
\frac{\partial \mathbf{H}}{\partial t}
\label{eq:7}
\end{equation}%
in which the displacement current is omitted. We assume non--relativistic
expansion velocity of the plasma cloud. Here $\mathbf{v}_{i}$ is the ion
mean velocity which is defined as
\begin{equation}
n_{e}=\sum_{\alpha} Z_{\alpha}n_{\alpha} , \quad \mathbf{v}_{i}
=\frac{1}{n_{e}}\sum_{\alpha} Z_{\alpha}n_{\alpha} \mathbf{v}_{\alpha} ,
\label{eq:8}
\end{equation}%
\begin{equation}
n_{\alpha}=\int f_{\alpha}d\mathbf{v} , \quad \mathbf{v}_{\alpha} =
\frac{1}{n_{\alpha}}\int \mathbf{v} f_{\alpha}d\mathbf{v} .
\label{eq:9}
\end{equation}%
The mean electron velocity $\mathbf{v}_{e}$ satisfies the equation
\begin{equation}
\mathbf{E}=-\frac{1}{c}\left[ \mathbf{v}_{e}\times \mathbf{H}\right]
\label{eq:5}
\end{equation}%
which can be derived from the hydrodynamic equation of motion of electrons
ignoring the inertial term (i.e. formally by taking the limit $m_{e} \to 0$).
The electric field in Eq.~\eqref{eq:5} is that required to keep the electrons
and ions together ensuring the quasineutrality of the plasma. Thus,
we do not resolve the expansion process on the Debye scale.
One can interpret Eq.~\eqref{eq:5} as follows. The electromagnetic
force acting on the electrons and ions does not depend on the particle
mass; therefore light electrons are accelerated much more than
ions. This would lead to charge separation and would violate the
quasineutrality condition if the strong Coulomb interaction would
not prevent this from occurring; i.e., an appropriate electric field
arises to enforce quasineutrality. The mean force itself, which so
strongly accelerates the electron gas, has to disappear or, more exactly,
its electric component has to compensate the Lorentz force within
the accuracy of the order of $m_{e}/m_{\alpha}$. This fact is expressed
by the approximate relation~\eqref{eq:5}. Equations~\eqref{eq:4}--\eqref{eq:5}
constitute a closed system and are used to investigate the collisionless
plasma dynamics in an external magnetic field.

Note that in Eqs.~\eqref{eq:4}--\eqref{eq:5} we have also neglected the terms
associated with the finite conductivity (e.g. magnetic field diffusion and Joule
heating). It can be shown that this approximation is easily satisfied in a wide
range of parameters. In fact, for example, we can estimate the characteristic
diffusion time for the magnetic field, $\tau_{m}=4\pi a_{L}^{2}\sigma /c^{2}$,
where $\sigma $ is the conductivity of the plasma. Introducing the characteristic
retardation time of the plasma cloud $\tau _{s}=R_{g}/u$ we obtain
$\tau _{m}/\tau _{s}\sim M_{A}^{2}(\lambda_{ii}/R_{g})\gg 1$. Here $\lambda _{ii}$
is the mean free path for ion--ion Coulomb interaction.

These approximations constitute the hybrid collisionless plasma model employed
in this paper. It should be emphasized that earlier simulations and experiments
demonstrated the practical realizability of the magnetolaminar interaction model
\cite{zak03,zak88} discussed briefly in Sec.~\ref{sec:1}. In addition it can be
shown that Eqs.~\eqref{eq:4}--\eqref{eq:5} of the hybrid model have universal
nature. This universality can be demonstrated by normalizing
Eqs.~\eqref{eq:4}--\eqref{eq:5} and all the quantities with
the scales $n_{b0}$ (densities), $c/\omega_{pb}$ (lengths), $v_{A}$ (velocities)
$H_{0}$ (magnetic field), $(v_{A}/c)H_{0}$ (electric field). In these units the
time and the ionic distribution functions are scaled by the cyclotron period
$\Omega^{-1}_{b}$ and $n_{b0}/v^{3}_{A}$, respectively. Here $\omega _{pb}$,
$\Omega _{b}$ and $n_{b0}$ are the plasma and cyclotron frequencies and the
unperturbed (initial) density of the background ions, respectively. $v_{A}$ is
the Alfv\'{e}n velocity in the background plasma. Then it is seen that
Eqs.~\eqref{eq:4}--\eqref{eq:5} depend only on the dimensionless parameters
$Z_{c}/Z_{b}$ and $Z_{c}m_{b}/Z_{b}m_{c}$.

\begin{figure}[tbp]
\includegraphics[width=11cm]{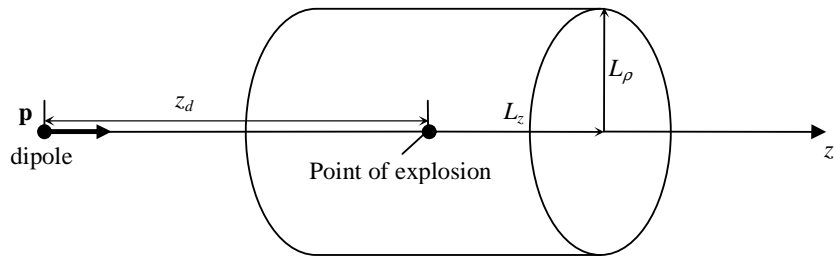}
\caption{Scheme of the simulation region showing location of the dipole
and the initial position of the plasma cloud.}
\label{fig:sch}
\end{figure}

Let us consider axially symmetric model for the dynamics of a point explosion
that forms a cloud of dense plasma expanding into a magnetized low--density
background plasma. For a simplicity we assume hydrogen cloud and background
plasmas with $m_{c}=m_{b}=m_{p}$, $Z_{c}=Z_{b}=1$, where $m_{p}$ is the proton
mass. At the initial time $t=0$, an explosion occurs at the point $\rho=z=0$
in a cylindrical region $0\leqslant \rho\leqslant L_{\rho}$,
$-L_{z}\leqslant z\leqslant L_{z}$ (see Fig.~\ref{fig:sch}) filled with a
homogeneous background plasma of density $n_{b0}$. The magnetic dipole with
a moment $\mathbf{p}=p_{z} \mathbf{e}_{z}$ creates a magnetic field
$\mathbf{H}_{0}(\mathbf{r})=\mathbf{H}_{0}(\boldsymbol{\rho},z)$ and is placed
outside the cylindrical region on the $z$--axis ($z_{d}<-L_{z}$,
$\boldsymbol{\rho}_{d}=0$). Here $\rho,z,\varphi$ are the cylindrical coordinates
and $\boldsymbol{\rho}$ is the cylindrical radius vector. Similarly $r,\theta,\varphi$
are the spherical coordinates with the radius vector $\mathbf{r}$, and $\theta$
is the angle between $\mathbf{r}$ and the $z$--axis. We consider the case when the
dipole moment is parallel to the $z$--axis and due to the symmetry we chose
$p_{z}>0$. In our simulations the dipole magnetic field is determined by the ratio
$\eta =H_{0}(0,-L_{z})/H_{0}(0,L_{z})>1$ and the strength of the dipole magnetic
field at the center of the simulation domain, i.e. $H_{0}=H_{0z}(0)>0$. Using these
two quantities it is easy to obtain the position $z_{d}$ and the moment $p_{z}$
of the dipole
\begin{equation}
p_{z}=-\frac{ z_{d} ^{3}}{2} H_{0} ,  \qquad
z_{d}=-L_{z}\frac{\eta^{1/3}+1}{\eta^{1/3}-1}
\label{eq:dip1}
\end{equation}
which completely determine the dipole field
for given length $L_{z}$. At fixed $H_{0}$ the limits $\eta\to 1$ and $\eta\gg 1$
correspond to nearly homogeneous and strongly inhomogeneous dipole magnetic fields
in a cylindrical region, respectively. Throughout this paper the cyclotron frequency
$\Omega_{b}$ and the Alfv\'{e}n velocity $v_{A}$ are determined in terms of the
magnetic field $H_{0}$. Note that due to the axial symmetry all physical quantities
are independent on the azimuthal angle $\varphi$ (obviously this symmetry breaks down
for arbitrary orientation of the magnetic moment $\mathbf{p}$). Here we neglect the
$\varphi $--derivatives in Eqs.~\eqref{eq:4}-\eqref{eq:5} and consider 2D3V model for
our simulations. It should be also emphasized that
the scheme outlined in Fig.~\ref{fig:sch} essentially differs from one adopted
in Refs.~\cite{win05,tan07} where the dipole is located inside the simulation
domain and due to the strong interactions the ions may be trapped by the dipole.

The explosion forms a cloud of dense plasma of radius $r_{0}$ containing $N_{c}$
ions with a total kinetic energy $W_{0}$. At the initial time, the velocity of
the cloud ions is distributed linearly along the radius, i.e. at $r\leqslant r_{0}$
\begin{equation}
v_{r}\left( r,0\right)=u(r) =v_{m}\frac{r}{r_{0}}
\label{eq:13}
\end{equation}%
and $v_{r}(r,0)=0$ at $r>r_{0}$. Here $v_{m}$ is the maximal velocity of the ions
in the cloud and is determined by the initial energy $W_{0}$ of the cloud, with
$v_{m}=(10W_{0}/3N_{c}m_{p})^{1/2}=v_{0}(5/3)^{1/2}$, where $v_{0}=(2W_{0}/N_{c}m_{p})^{1/2}$
is the average velocity of the cloud ions.

The equation of the energy conservation of the system is given by
\begin{equation}
W =W_{\mathrm{kin}}+2\pi \int_{0}^{L_{\rho}}\rho d\rho
\int_{-L_{z}}^{L_{z}} dz \left( \frac{1}{2%
}nmv_{e}^{2}+\frac{H^{2}}{8\pi }\right) .
\label{eq:14}
\end{equation}
Here $n=n_{e}$, and $n$ is the total density of the ions in the cloud and
background plasmas. The total energy of the system consists of the kinetic
energies of the electron gas, the energy of the magnetic field, and the
kinetic energy $W_{\mathrm{kin}}$ of the ions. Note that in Eq.~\eqref{eq:14}
we have neglected the energy of the electric field as $v_{e}\ll c$ and $E\ll H$.

The system size employed in the computations is $2L_{z}=9.8a_{L}$ and $L_{\rho}=4.9a_{L}$
and uses $98\times 49$ cells in $z$ and $\rho $ directions, respectively, where
$a_{L}=v_{m}/\Omega_{b}$ is the ion cyclotron radius. Thus, the cell size in the
$\rho $ and $z$ directions is $h=h_{\rho}=h_{z}=0.1a_{L}$. In order to correctly
resolve the ion gyromotion the quantity $h$ and the time step $\Delta t$ are fraction
of the ion cyclotron radius $a_{L}$ and the ion cyclotron period $\Omega^{-1}_{b}$,
respectively. In our simulations $\Delta t=0.01\Omega^{-1}_{b}$. The initial
radius $r_{0}$ of the plasma cloud is considerably smaller than the step size $h$;
that is, we can assume that at the initial time the
cloud is concentrated at a point. At the initial time $t=0$ we take the background
plasma at rest and assume that the background particles are distributed uniformly
over the entire cylindrical region and the particles of the expanding plasma are
distributed uniformly in the cloud. Explicitly the initial distribution functions
of the background and cloud ions are given by $f_{b}(\mathbf{r},\mathbf{v},0)= %
n_{b0}\delta (\mathbf{v})$ and $f_{c}(\mathbf{r},\mathbf{v},0) =n_{c0}\delta %
(\mathbf{v}-u(r)\frac{\mathbf{r}}{r})$, respectively, where $n_{c0}$ is the initial
density of the cloud ions and $u(r)$ is determined by Eq.~\eqref{eq:13}. Note that
the distribution function $f_{c}(\mathbf{r},\mathbf{v},0)$ is non--zero only at
$0\leqslant r\leqslant r_0$. Boundary conditions are one of the most important
conditions to be satisfied in an electromagnetic problem, and they vary depending
on the problem. In Ref.~\cite{win05}, periodic boundary conditions are imposed on
the fields and the particles. They only run the system a relatively short amount
of time before any disturbances reach the edges of the system. In this paper, at
the boundaries of the cylindrical region $\rho=L_{\rho}$ and $z=\pm L_{z}$, all
quantities are specified by its unperturbed values, and on the $z$--axis (i.e. at
$\rho=0$) we assume the condition $E_{\rho ,\varphi}=H_{\rho ,\varphi}=v_{e\rho , %
\varphi}=0$ which is a natural consequence of the symmetry of the model.
The absorption boundary conditions are chosen for all the particles which means that
any particle leaving the computation domain is assumed to have gotten lost.
With these boundary conditions the calculations are continued until the time
the perturbations reach the boundaries of the cylindrical region.

\begin{table*}[tbp]
\caption{The values of the parameters for the numerical simulation
($M_{A}=15$, $\delta =1$). The number of the ions (protons) in a
cloud is $N_{c}=1.67\times 10^{19}$.}
\label{tab:1}
\begin{tabular}{llll}\hline\hline
$a_{L}=34.16$ cm & $R_{g}=34.16$ cm &  $R_{m}=175.2$ cm  &  $\tau _{s}=1.044$ $\mu$s  \\
\hline
$H_{0}=100$ G & $W_{0}=0.896$ kJ & $v_{m}=3.27\times 10^{7}$ cm/s & $\eta =100$  \\
\hline
$n_{b0}=10^{14}$ cm$^{-3}$ & $v_{A}=2.18\times 10^{6}$ cm/s & $%
\Omega _{b}=9.58 \times 10^{5}$ s$^{-1}$ & $\omega
_{pb}=1.32\times 10^{10}$ s$^{-1}$ \\
\hline
\end{tabular}
\end{table*}

The equations of motion for the ions are the equations of the characteristics
of the Vlasov kinetic equation \eqref{eq:4}. Assuming hydrogen plasma we obtain
\begin{equation}
\dot{\mathbf{r}}_{i} ={\mathbf{v}}_{i}, \qquad \dot{\mathbf{v}}_{i} =%
\frac{e}{m_{p}}\left( {\mathbf{E}}+\frac{1}{c}\left[
{\mathbf{v}}_{i}\times {\mathbf{H}}\right] \right) .
\label{eq:17}
\end{equation}
These equations are solved by the particle-in-cell (PIC) method using Boris pusher
algorithm~\cite{hoc88,bir04}. Instead of the kinetic equation integration, the
trajectories of a large number of the quasi--particles (each of these particles in
turn consist of the large number of ions) are computed. In our simulations the total
number of the quasi--particles is $97000$ and $20168$ quasi--particles are used for
a cloud. For the background plasma we set $16$ quasi--particles in the cell with close
values of the cyclotron radii (because $h\ll a_{L}$) which allows us to simulate ion
kinetics in details. Further increasing the number of the quasi--particles does not
essentially influence the simulation results. The Maxwell equations are solved using
an explicit first order splitting scheme on staggered grid. A more detailed description
of the mathematical model and the numerical simulation can be found, for example, in
Refs.~\cite{ler84,hoc88,bir04}.

\section{Results}
\label{sec:3}

In this section, results from several calculations are presented
to describe the overall physics and to show the properties of the
resulting structures. Numerical simulations have been performed for
calculation of the energy, density and electromagnetic characteristics of
the expansion of a hydrogen plasma into a uniform magnetized background
hydrogen plasma at high Alfv\'{e}n--Mach numbers with $M_{A}=v_{m}/v_{A}=15.0$
and for a MLM interaction parameter $\delta =1.0$. The basic physical
parameters of the simulations are shown in Table~\ref{tab:1}, where $\tau_{s}=R_{g}/v_{m}$
is the hydrodynamic retardation time. Note that $R_{g}\ll R_{m}$ and
the retardation of the cloud is caused here by the interaction with the background plasma.
The dipole is placed at $z_{d}=-258.86$~cm (see Fig.~\ref{fig:sch}).
The initial radius of the plasma cloud is $r_{0}=0.01a_{L}$ which corresponds
to the initial density $n_{c0}= 10^{20}$~cm$^{-3}$. The parameter $\kappa$
introduced in Sec.~\ref{sec:1} is $\kappa=0.62>\kappa_{c}$. Thus we expect
that the ions will not be captured by a dipole magnetic field.
Using the parameters introduced above, the simulation is run to $t\simeq %
5\tau_{s}$ when the perturbation reach to the boundaries of the cylindrical
region and the boundary conditions adopted here become invalid. Due to the
cylindrical symmetry the solutions of Eqs.~\eqref{eq:4}--\eqref{eq:5} are
obtained using the cylindrical coordinate system shown in Fig.~\ref{fig:sch}.

\begin{figure}[tbp]
\includegraphics[width=16cm]{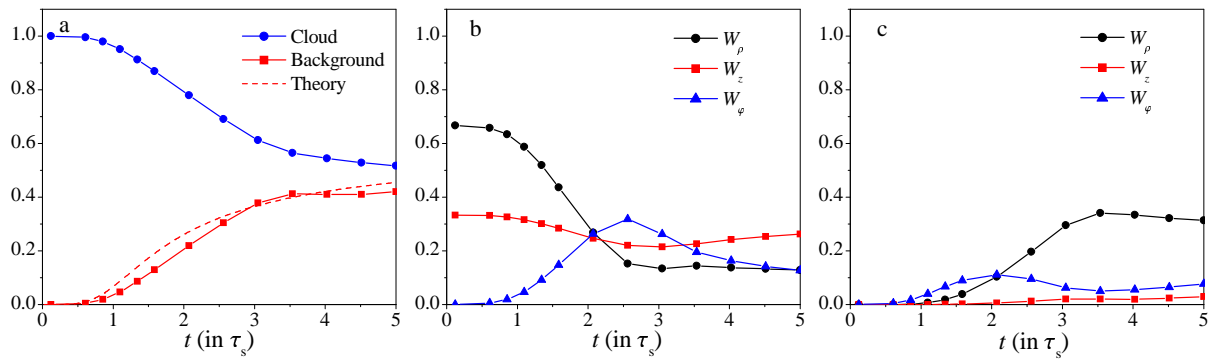}
\caption{(Color online) The time (in units of $\tau_{s}$) variation of the kinetic energy
(in units of $W_{0}$) of the plasma cloud and background plasma particles
for $M_{A}=15.0$, $\delta
=1.0$. a) The variation of the total energy of the background
plasma (squares) and plasma cloud (circles). The dashed line is the theoretical
predictions of Ref.~\cite{gol78} (see the text for details).
b) The time variation of the
radial ($W_{\rho}$, circles), longitudinal ($W_{z}$, squares) and rotational
($W_{\varphi}$, triangles) energies of the plasma cloud. c) The same as in b) but
for background plasma.}
\label{fig:1}
\end{figure}

Figure~\ref{fig:1} shows the time variation of the kinetic energies of
the cloud and the background plasma. The energies are normalized to the
initial energy of the plasma cloud $W_{0}$ and the time is given in units
of the retardation time $\tau_{s}$. It is seen that the energy of the plasma
cloud decreases with a time and at $t\simeq 5\tau _{s}$, about a half of the initial
energy is transferred to the background plasma. Let us recall that the
Coulomb collisions between particles are completely ignored here and thus
as mentioned in Sec.~\ref{sec:1} the energy transfer is caused by the
MLM interaction. In Ref.~\cite{gol78} a simple analytical model was suggested
for the energy transfer from the plasma cloud to the ambient plasma in the
presence of a homogeneous magnetic field. Assuming the initial distribution
functions $f_{c}(\mathbf{r},\mathbf{v},0)=N_{c}\delta(\mathbf{r})f_{0}(v)$
and $f_{b}(\mathbf{r},\mathbf{v},0)=n_{b}\delta (\mathbf{v})$ of the ions in
the plasma cloud and the background plasma, respectively, the energy transfer
in this model reads (see Ref.~\cite{gol78} for details)
\begin{equation}
\frac{\Delta W(t)}{W_{0}}=\frac{\delta }{6}\left[\kappa ^{5}(t)
+5\int_{\kappa (t)} ^{\infty }\chi ^{2}\left( x,t
\right) dx\right] ,
\label{eq:ch1}
\end{equation}
where
\begin{equation}
\varphi \left(u \right) =4\pi \int _{u}^{\infty} f_{0}(v) v^{2} dv, \quad
\chi \left( x,t \right) =x^{2}-\left[ x^{3} -\varphi \left(\frac{R_{g} x}{t} \right)\right] ^{2/3} .
\label{eq:ch2}
\end{equation}
Here the velocity distribution function $f_{0}(v)$ is normalized according to
$\varphi (0)=1$ and the quantity $\kappa (t)$ is determined from the equation
\begin{equation}
\kappa ^{3}(t) =\varphi \left(\frac{R_{g}\kappa (t)}{t}\right) .
\label{eq:ch3}
\end{equation}
Assuming that the strength of the homogeneous magnetic field is $H_{0}$ and the
velocity distribution function $f_{0}(v) =(3/4\pi v^{3}_{m})H(v_{m}-v)$ (where
$H(z)$ is the Heaviside unit--step function) the prediction of this model is shown
in Fig.~\ref{fig:1}a (the dashed line). The transferred energy \eqref{eq:ch1} at
$t<\tau_{s}$ and for a chosen $f_{0}(v)$ is then given by $\Delta W(t)/W_{0}\simeq (\delta /6)(t/\tau_{s})^{5}$
and in the limit $t\to \infty$ approaches $\Delta W_{\infty}/W_{0}=C\delta$ (where
$C\simeq 0.59$) independently of the particles initial velocity distribution
function in a cloud. It is seen that the theory agrees qualitatively with the
simulation (squares) although it has been derived under assumption that
$\Delta W(t)\ll W_{0}$. The sum of the cloud and background plasma energies is
the total kinetic energy $W_{\mathrm{kin}}(t)$ of the ions involved in the energy
conservation relation, Eq.~\eqref{eq:14}. This energy decreases with a time due
to the energy transfer to the electrons (second term in Eq.~\eqref{eq:14}) and
the magnetic field (last term in Eq.~\eqref{eq:14}). However the amount of the
energy gained by the electrons and the magnetic field is small compared to
$W_{\mathrm{kin}}$.

The time variation of the radial ($W_{\rho}$), longitudinal ($W_{z}$) and
gyromotion ($W_{\varphi}$) components of the energies are shown in
Fig.~\ref{fig:1}b,c. Using Eq.~\eqref{eq:13} it is easy to calculate the radial
and longitudinal energies of the plasma cloud at $t=0$ which are given by
$W_{\rho}(0)=(2/3)W_{0}$ and $W_{z}(0)=(1/3)W_{0}$, respectively. At the initial
time there is no gyromotion of the particles and $W_{\varphi}(0)=0$. As
shown in Fig.~\ref{fig:1}b,c the expanding plasma cloud loses its radial energy
which is transferred to the gyromotion energy of the cloud and the background
plasma ions. In addition a large amount of the cloud kinetic energy is
transferred to the radial energy of the background ions. It is also seen that
the changes of the longitudinal energies of the ions are small.

\begin{figure}[tbp]
\includegraphics[width=16cm]{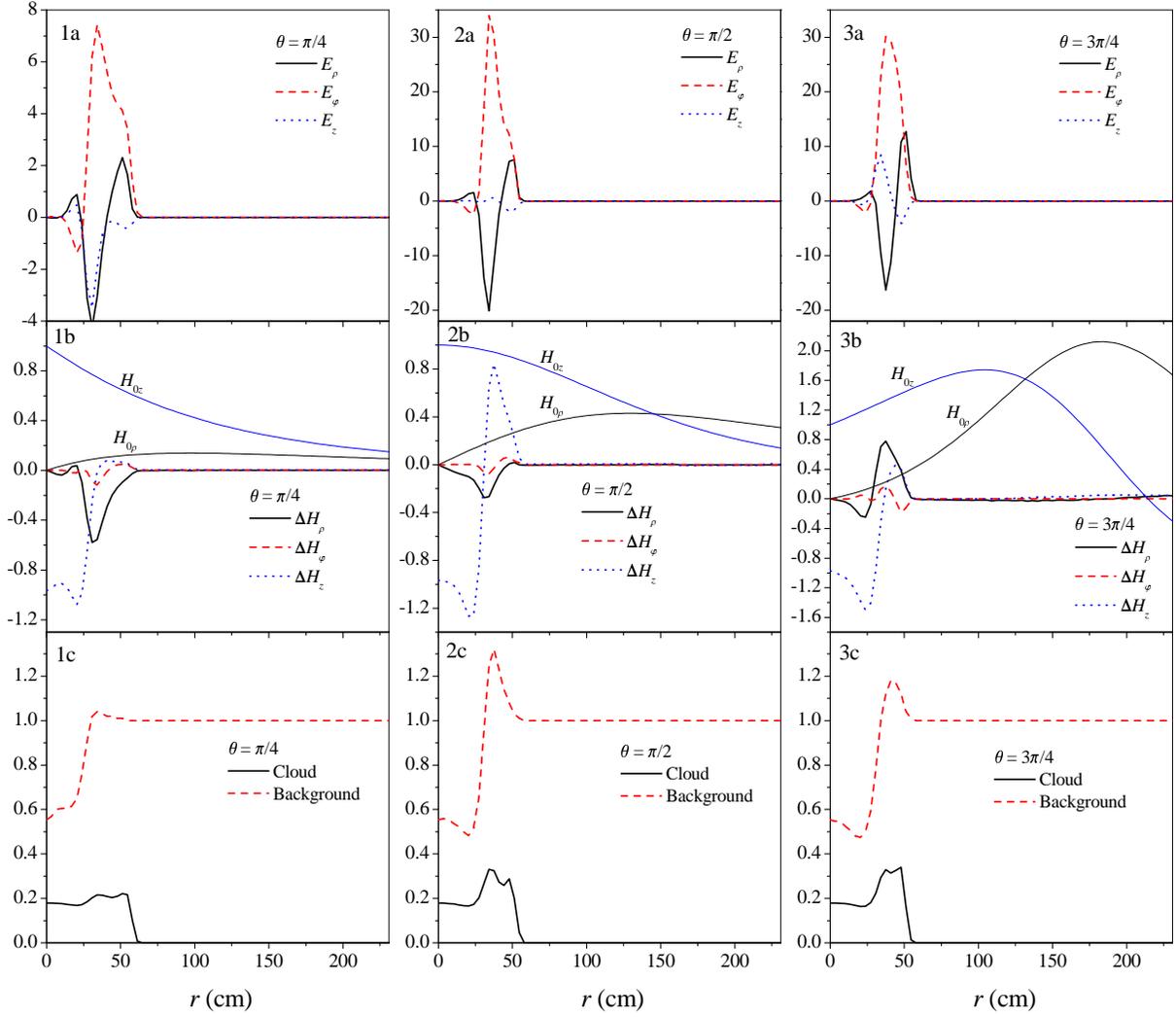}
\caption{(Color online) The perturbations of the electromagnetic fields and the densities of
the plasma cloud and background plasma as functions of the spherical
coordinate $r$ (in cm) for $t=2\tau _{s}$, $M_{A}=15.0$ and $\delta =1.0$
and for some values of $\theta$. The panels
1a--3a show the distributions of the $\rho$ (solid lines), $\varphi $
(dashed lines), and $z$ (dotted lines) components of the electric field
(in units of $(v_{A}/c)H_{0}$) at $\theta =\pi/4$, $\theta =\pi/2$ and
$\theta =3\pi/4$, respectively. The panels 1b--3b show the same as 1a--3a
but for the perturbation of the magnetic field $\Delta \mathbf{H}$ (in units of $H_{0}$).
The components of the dipole magnetic field $H_{0z}$ and $H_{0\rho}$ are
also shown as thin solid lines. 1c--3c show the densities
(in units of $n_{b0}$) of the cloud (solid lines) and the background plasma
(dashed lines) at $\theta =\pi/4$, $\theta =\pi/2$ and $\theta =3\pi/4$,
respectively.}
\label{fig:2}
\end{figure}

Figure~\ref{fig:2} shows the electric (panels 1a--3a) and the perturbed magnetic
($\Delta \mathbf{H}(\mathbf{r},t)=\mathbf{H}(\mathbf{r},t)-\mathbf{H}_{0}(\mathbf{r})$,
panels 1b--3b) fields strengths in units of $(v_{A}/c)H_{0}$ and $H_{0}$, respectively,
as well as the densities (panels 1c--3c) of the cloud and the background
plasma in units of $n_{b0}$ as the functions of $r$ (in cm) at $t=2\tau _{s}$,
$\theta =\pi /4$, $\theta =\pi /2$ and $\theta =3\pi /4$. Here $\theta$ is
the angle between the spherical radius vector $\mathbf{r}$ and the $z$--axis. In
the panels 1b--3b of Fig.~\ref{fig:2} the radial $H_{0\rho}$ and longitudinal
$H_{0z}$ components of the dipole magnetic field are also shown (note that
$H_{0\varphi}=0$) as thin solid lines. As expected the density of the plasma
cloud is strongly reduced due to the expansion. In addition the perturbations of the
electromagnetic fields are larger at $\theta \geqslant \pi/2$ where the dipole
magnetic field is stronger. At the initial stage of the expansion process
($t\ll \tau_{s}$) the ions of the plasma cloud front propagates through a quiescent
background plasma, and different plasmas are mixed. At this stage the background
ions are accelerated mainly by the generated azimuthal electric field. Then,
turning under the Lorentz force action, they acquire the radial velocity. At
the later stage when the radial and longitudinal electric fields arise, the
motionless background ions acquire a velocity due to all components of the
field, with the azimuthal and radial fields being dominant. In general
the motion of the ions in the electric and magnetic fields is very complicated. They
will do a gyromotion along the magnetic field lines. But along with the gyromotion,
the ions will also experience $\mathbf{E}\times \mathbf{H}$ drift and gradient drift.
In contrast to the azimuthal electric field, the radial and longitudinal components
of $\mathbf{E}$ are space--charge fields. This follows from Eq.~\eqref{eq:5} as
the electromagnetic induction Eq.~\eqref{eq:7} provides the azimuthal field generation
rather than the radial and longitudinal ones. The electric field is generated in a
region of the perturbation of the magnetic field, and everywhere it is perpendicular
to the total field $\mathbf{H}$.

\begin{figure}[tbp]
\includegraphics[width=9cm]{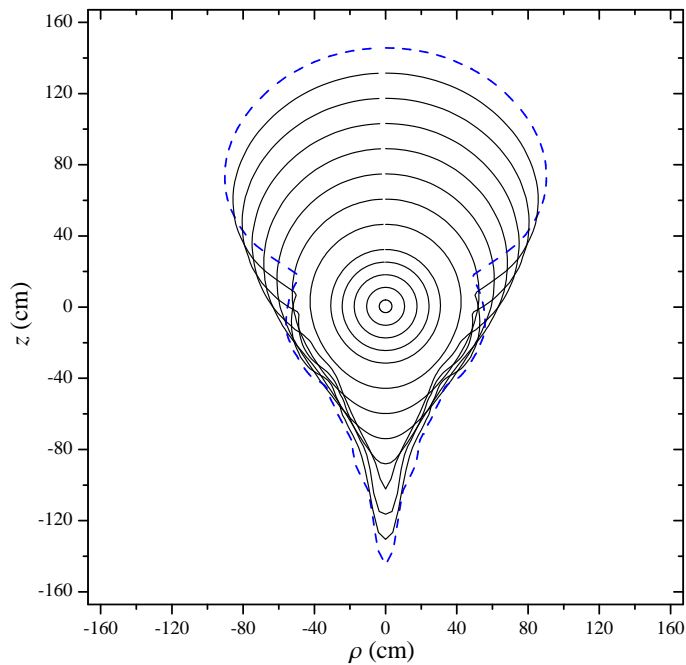}
\caption{(Color online) The evolution of the boundary of the plasma cloud from $t=0.11\tau_{s}$ (central
circle) to $t\simeq 5\tau_{s}$ (external dashed line). The time interval between two
successive lines is approximately $\Delta t\simeq 0.4\tau_{s}$.}
\label{fig:3}
\end{figure}

From Fig.~\ref{fig:2} it is seen that the retardation of the plasma cloud is
accompanied by the formation of a compressed plasma layer which moves together with
the compressed electromagnetic field; i.e., a collisionless shock wave is formed.
The depth of the compressed layer is of the order of the ion cyclotron radius
$a_{L}$, in agreement with an estimate $\Delta \sim a_{L}$ of the width $\Delta$
of the collisionless shock front given in Ref.~\cite{sag66}. The ions of the background
plasma are pushed out of the expansion region; this leads to the formation of a plasma
cavity. It correlates with the diamagnetic cavity, a region in which the total magnetic
field  $\mathbf{H}(\mathbf{r},t)$ is smaller than the unperturbed dipole field because
it is squeezed out (Fig.~\ref{fig:2}). Thus, there is essentially no electric field in
the cavity. Due to the retardation of the ions of the plasma cloud front, a part of its
mass forms a shell near the front boundary, see panels 1c--3c of Fig.~\ref{fig:2}.
However this shell is not isotropic and is more
visible at $\theta\geqslant \pi/2$, i.e. in the region closer to the dipole. At the
later time of the expansion $t>2\tau_{s}$ the plasma shell becomes more and more
pronounced with the formation of the cavity also in the plasma cloud with sharply
(with the width $\sim a_{L}$) distributed ions.

The time evolution of the plasma cloud boundary is shown in Fig.~\ref{fig:3}, where
the central circle and external dashed lines correspond to $t=0.11\tau_{s}$ and $t\simeq 5\tau_{s}$,
respectively. As shown in Fig.~\ref{fig:3} an initially symmetrical plasma cloud
begins to be asymmetric in the $\rho z$ plane at $t\gtrsim 2\tau_{s}$. On the other
hand the shape of the cloud is symmetrical at all the stages in the plane perpendicular
to the $z$--axis (not shown). The similar shape of the plasma cloud is observed
in the experiments as well as in numerical simulations of the plasma expansion in a
vacuum when the background plasma is absent \cite{zak03,mur01}. In this case the
asymmetrical pattern may be caused only by the ambient magnetic
field. At the initial stage, the plasma cloud expands isotropically as a free stream
with high kinetic energy. Then, plasma expanding downward begins to decelerate because
the ambient dipole magnetic field is stronger at a lower region near the dipole
(see, e.g., Figs.~\ref{fig:sch} and \ref{fig:3}). The kinetic beta $\beta_{c}$ defined as
the ratio of plasma cloud kinetic energy density to magnetic energy density reaches unity
around this area first, and then strong interaction between plasma and the ambient field
occurs. Thus, the diamagnetic motion of plasma is induced to generate the surface
diamagnetic current asymmetrically, only on the lower plasma surface. In Ref.~\cite{ner06}
an analytical model was suggested to calculate the pressure
of the dipole magnetic field on the surface of the plasma cloud. It was shown
that at $\mathbf{p}=p_{z}\mathbf{e}_{z}$ this pressure vanishes at $\theta =0;\pi$ and
has its maximum at $\theta =\theta_{\max}$
($\pi/2 <\theta_{\max} <\pi$) which depends on the ratio $R/|z_{d}|$. Here $R$ is the radius of the
plasma cloud. The maximal value $\theta_{\max}$ tends to $\pi$ with increasing $R$
and shifts towards the value $\theta_{\max}\simeq \pi/2$, when $R\to 0$. Therefore
the layer near $\theta\simeq \theta_{\max}$
of the expanding plasma sphere will be mainly deformed by the external magnetic pressure.
In the present context of the high--energy expansion $\beta_{c}\gg 1$ even at the later
stages of the expansion and the magnetic field cannot deform directly the cloud shape.
In this case the deformation occurs due to the interaction with the background plasma
with initially low kinetic $\beta_{b}$. Then the background plasma tends to change
the shape to follow the dipole magnetic field lines. As shown in Fig.~\ref{fig:1} the kinetic
energy of the background ions and hence the parameter $\beta_{b}$ increases with time
which leads to the formation of the asymmetrical kinetic pressure of the background ions
on the plasma cloud.

\section{Conclusions}
\label{sec:conc}

In this paper we have investigated the collisionless expansion of the dense plasma
cloud into magnetized background plasma in the presence of a dipole magnetic field.
The 2D3V hybrid--PIC simulations with kinetic ions and massless fluid electrons
have been employed. We have considered the case of high--energy super--Alfv\'{e}nic
expansion with $M_{A}\gg 1$ when the expanding plasma is not captured by the ambient
magnetic field. It is shown that in this parameter regime the retardation and the
deformation of the plasma cloud is mainly caused by the interaction with background
plasma which, however, can be strongly affected (at least at the initial stage) by
a magnetic field. Our numerical results for the energy transfer from the plasma cloud
to the background plasma have been compared with the theoretical predictions of
Ref.~\cite{gol78} and qualitatively good agreement has been found.
For future applications the arbitrary orientation of the dipole (with respect to the
$z$--axis, see Fig.~\ref{fig:sch}) should be considered. For instance, such a
configuration has been realized in the experiment \cite{mur01}. In this case the
physical quantities also depend on the azimuthal angle $\varphi$ and the shape of the
plasma cloud becomes asymmetric in the plane perpendicular to the $z$--axis.

An additional item for further investigations is the development of a model which
accounts the thermal effects for electrons which are completely neglected in the
current study. These effects can be included by adding in the right hand side of
Eq.~\eqref{eq:5} the term $-(1/en)\boldsymbol{\nabla} p_{e}$, where $p_{e}$ is the
thermal pressure of the electrons. Also along with Eqs.~\eqref{eq:4}--\eqref{eq:5}
an equation for the evolution of $p_{e}$ should be considered (see, e.g., Ref.~\cite{osi03}).
In this connection assuming the strong heating of the electrons in the shock front
it is also desirable to investigate the physics allowing for charge separation
effects and the resolution of the electron Debye length scale. We intend to address
this and other issues in the context of the plasma expansion in a separate study.

\begin{acknowledgments}
This work has been partially supported by the Armenian Ministry
of Higher Education and Science (Project No.~87).
\end{acknowledgments}

\end{document}